\documentclass[conference]{IEEEtran}
\IEEEoverridecommandlockouts

\usepackage{cite}
\usepackage{amsmath,amssymb,amsfonts}
\usepackage{booktabs}
\usepackage{graphicx}
\usepackage{textcomp}
\usepackage{url}

\graphicspath{{figures/}}

\begin{document}

\title{Label Authority in Speech Emotion Recognition: Intended and Perceived Ground Truth in Emozionalmente%
\thanks{M. Giordano, S. Giacomelli and F. Graziosi are with the Department of
Information Engineering, Computer Science and Mathematics (DISIM), University
of L'Aquila, L'Aquila, 67100, ITA
(stefano.giacomelli@graduate.univaq.it,
marco.giordano3@graduate.univaq.it, fabio.graziosi@univaq.it). C. Rinaldi is
with the National Inter-University Consortium for Telecommunications (CNIT),
University of L'Aquila, L'Aquila, 67100, ITA
(claudia.rinaldi@univaq.it). This work was partially supported by the European
Union under the Italian National Recovery and Resilience Plan (NRRP) of
NextGenerationEU, partnership on ``Telecommunications of the Future''
(PE00000001--program ``RESTART'').}}

\author{
\IEEEauthorblockN{Marco Giordano}
\IEEEauthorblockA{\textit{Student Member, IEEE}\\}
\and
\IEEEauthorblockN{Stefano Giacomelli}
\IEEEauthorblockA{\textit{Graduate Student Member, IEEE}\\}
\and
\IEEEauthorblockN{Claudia Rinaldi}
\IEEEauthorblockA{\textit{Member, IEEE}\\}
\and
\IEEEauthorblockN{Fabio Graziosi}
\IEEEauthorblockA{\textit{Member, IEEE}\\}
}

\maketitle

\begin{abstract}
Ground truth in Speech Emotion Recognition (SER) is often presented as a
single categorical fact, yet an acted emotion and the emotion heard by
listeners describe different aspects of the same recording. We examine this
distinction in Emozionalmente, an acted Italian corpus that provides both an
elicited production label and five listener judgements for every clip. Using
frozen XLS-R, HuBERT, and WavLM representations with matched probes, we compare
three label authorities: the expressed condition, the listeners' hard
plurality, and their complete vote distribution.

The results tell a consistent story. Recovering the expressed condition is
easier than predicting a tie-broken listener plurality, but that gap compares
two different questions rather than revealing a general loss of SER ability.
As listener agreement weakens, the plurality becomes an increasingly brittle
hard target. Training on the full vote distribution does not substantially
change which class wins the model's hard decision, but it produces probability
assignments that more closely follow the alternatives selected by listeners,
especially for ambiguous clips.

No authority therefore replaces the others. Expressed labels describe how
speech was produced, pluralities provide an operational single-class decision,
and vote distributions preserve perceptual uncertainty. For human-facing
Internet of Sounds applications, evaluations should state which authority is
being predicted, retain its aggregation rule and agreement information, and
use distributional scores when disagreement is part of the target.
\end{abstract}


\section{Introduction}

Speech Emotion Recognition (SER) is usually formulated as classification: an
utterance enters a model and one emotion label comes out. The apparent
simplicity of that formulation hides a prior decision about what the label is
supposed to mean. In acted corpora, it commonly records the emotion requested
during data collection. Such a label is a useful and reproducible description
of the production protocol, but it is not a direct observation of what a
listener heard, and still less of what a speaker privately felt. Conversely, a
listener label records a perceptual judgement whose meaning depends on who
listened, under which conditions, and how several judgements were aggregated.
Expression--perception mismatches have been documented even in widely used
emotional-speech resources \cite{busso2008expressionPerception}. Calling both
sources ``ground truth'' without qualification therefore makes two different
questions look like one.

This distinction matters whenever an affective audio model becomes part of a
human-facing Internet of Sounds application. A semantic archive may index how
content was deliberately performed, while a search or accessibility interface
may need to retrieve how that content is likely to be perceived. An adaptive
service may need a categorical decision, yet also need to know when listeners
would plausibly disagree before changing its behaviour. These uses do not
require the same supervisory authority. A system can correctly recover an
elicited production condition while disagreeing with its users, or match a
listener plurality while concealing that the plurality was decided by a tie.

We call this choice the \emph{label-authority problem}. It concerns which
observation is authorized to define success, rather than whether one annotation
file is cleaner than another. We investigate the problem in Emozionalmente, a
crowdsourced Italian corpus of simulated emotional speech that provides an
elicited production label and five independent post-hoc recognition judgements
for every clip \cite{catania2025emozionalmente}. The corpus makes it possible to
study three operational authorities on exactly the same audio: the expressed
condition, a single recognized class obtained by plurality, and the complete
distribution of listener votes. None is assumed to be universally correct.
Each describes a different aspect of the recording and induces a different
evaluation question.

The study follows that distinction from people to models. It first asks how the
expressed condition and listener responses diverge before any classifier is
introduced. It then compares matched probes on frozen XLS-R, HuBERT, and WavLM
representations, separating changes in the evaluation reference from changes
in the training target. Finally, it asks whether distributional supervision
retains perceptual alternatives that disappear when five votes are collapsed
to one tie-broken class. Run-to-run variation, paired clip bootstraps,
agreement-stratified analyses, and configuration-matched sensitivities are used
for different uncertainty questions; the many layer conditions are not treated
as independent experimental replicates.

The resulting picture is more informative than a ranking of three targets.
Recovering the expressed condition is easier than predicting a listener
plurality, but the corresponding scores refer to different tasks. The plurality
becomes increasingly brittle as agreement weakens. Soft supervision does not
materially change hard plurality UAR on the full test split, yet it gives
probability distributions that are substantially closer to the complete vote
histograms, especially on ambiguous clips. Model confusions also concentrate
on the same emotion pairs that human listeners confuse most often. Together,
these observations support a task-dependent reporting rule: name the label
authority, preserve its aggregation and agreement information, and use a
proper distributional score when disagreement is part of the phenomenon being
modelled.

The contribution is methodological rather than architectural. The evidence is
restricted to one acted Italian corpus and to frozen representations with
shallow probes. Within that scope, the paper contributes a human-first account
of the three authorities, a matched hard and distributional evaluation, an
analysis of how model errors track perceptual ambiguity, and concrete guidance
for specifying and reporting label provenance in human-facing audio systems.

\section{Background}

\subsection{Authority Is Not Reliability}

SER surveys describe a field in which corpus design, emotion representation,
and evaluation protocol vary substantially \cite{latif2023serRepresentationSurvey,
deLope2023serReview}. Acted resources remain useful because their elicitation
procedures provide balanced and interpretable production conditions. Italian
corpora such as EMOVO and DEMoS, for example, deliberately record performed
emotions under controlled protocols \cite{costantini2014emovo,
paradaCabaleiro2020demos}. IEMOCAP combines elicited interaction with several
forms of annotation and thereby illustrates how one recording can support
multiple descriptions \cite{busso2008iemocap}. The methodological issue is not
that an acted label is false. It is that a production label, a perception label,
and a felt-emotion report are observations of different constructs.

This is why label authority must be separated from label reliability.
Reliability coefficients describe how consistently annotators apply a scheme
\cite{artstein2017interannotator}. They do not decide whether a study should
predict the elicited condition, a listener consensus, or an individual
perception. A perfectly reliable panel could still answer a different question
from the production protocol; an imperfectly reliable panel can still provide
valuable evidence about perceptual ambiguity. The authority identifies the
estimand. Reliability characterizes the evidence available for that authority.

Perception is also conditional on the annotation procedure. Utterance order and
context alter crowdsourced emotion judgements \cite{jaiswal2019museOrdering},
while audio-only and multimodal evaluations need not support the same SER
conclusions \cite{chou2025stimulusModality}. These effects do not make listener
responses unusable. They make their provenance part of the measurement. Recent
work on EmoDB similarly argues that removing ambiguous samples can discard
scientifically useful structure rather than merely clean a corpus
\cite{burkhardt2025emodb}.

\subsection{From Consensus Labels to Distributions}

A plurality turns several judgements into the single class required by a
conventional classifier. This is operationally convenient, but the conversion
is lossy. Minority responses disappear, the strength of the winning class is
hidden, and ties require an external rule. The resulting hard target can make a
model look wrong even when it favours an alternative selected by equally many
listeners. Agreement statistics and plurality accuracy therefore answer
different questions: one describes the panel, while the other scores a model
against the panel's compressed outcome.

Several strands of SER research retain more of the annotation structure.
Curriculum and soft-label approaches use disagreement to order or smooth
training examples \cite{lotfian2019curriculum,fayek2016softLabels}. Label
distribution learning treats annotator proportions as the target itself, and
Bayesian or distributional formulations explicitly represent label uncertainty
\cite{prabhu2022tDistributions,prabhu2024labelUncertainty}. Work that models
individual raters, rather than only their aggregate, further shows that
subjective and consensus labels can contain complementary information
\cite{chou2019every}. These methods differ in architecture and statistical
interpretation, but they share an important premise: disagreement may be part
of the phenomenon rather than a nuisance to erase.

The present study adopts the simplest version of that premise. The five
listener votes are retained as an empirical categorical distribution, without
claiming that five observations reveal a latent population distribution
exactly. A hard plurality and a soft distribution can then be evaluated on
shared clips and against shared references. This controlled comparison is
particularly important because a higher hard score and a better probabilistic
fit need not coincide.

\subsection{Frozen Representations as a Diagnostic}

Self-supervised speech models provide a common representational substrate for
the comparison. XLS-R learns multilingual acoustic representations through a
wav2vec~2.0-style contrastive objective \cite{babu2021xlsr}; HuBERT predicts
cluster-derived hidden units \cite{hsu2021hubert}; and WavLM extends masked
speech modelling to noisy and overlapping conditions \cite{chen2022wavlm}.
Freezing these backbones and varying only shallow probes makes the experiment a
diagnostic of accessible information under matched conditions. It deliberately
does not answer how end-to-end fine-tuning, larger annotation panels, or a
different corpus would change the result.

\section{Data and Label Authorities}

\subsection{Corpus and Listener Votes}

Emozionalmente contains 6,902 utterances from 431 speakers and approximately
7.3 hours of mono, 16-kHz speech \cite{catania2025emozionalmente}. Its seven
categories are \textit{anger}, \textit{disgust}, \textit{fear}, \textit{joy},
\textit{neutrality}, \textit{sadness}, and \textit{surprise}. The elicitation
design is balanced, with 986 clips for each expressed category. After recording,
829 listeners supplied 34,510 valid recognition judgements: exactly five for
every clip. Each judgement also carries an audio-quality flag.

The availability of paired production and perception metadata is the reason
for using one corpus rather than combining several datasets. Cross-corpus
differences in speakers, acoustics, classes, and annotation would otherwise be
confounded with the authority being studied. All comparisons here therefore
refer to the same recordings and the same speaker-independent partitions.

\subsection{Three Operational Authorities}

The \emph{expressed} label records the emotion assigned to the speaker during
elicitation. It is encoded as a one-hot target and answers whether a model can
recover the controlled production condition. The \emph{recognized plurality}
is the most frequent of the five listener votes. Because a hard classifier
requires one class, tied pluralities are resolved deterministically by the
lowest configured class index. That convention makes the experiment
reproducible; it does not turn the selected class into a stronger consensus.

The \emph{recognized distribution} retains the empirical vote proportions. If
three listeners select joy and two select surprise, the target assigns 0.6 to
joy and 0.4 to surprise instead of replacing the panel with a one-hot joy
label. Formally, $p_c=n_c/5$, where $n_c$ is the number of votes for class $c$.
We retain the top-vote share, tie status, entropy, and quality flags alongside
these targets. Table~\ref{tab:authorities} summarizes the questions they make
measurable.

\begin{table*}[t]
\caption{Operational label authorities. ``Reference unit'' identifies what one
evaluation item represents; it is not a claim that one authority is universally
more truthful than another.}
\label{tab:authorities}
\centering
\footnotesize
\setlength{\tabcolsep}{5pt}
\begin{tabular}{@{}p{0.19\textwidth}p{0.23\textwidth}p{0.24\textwidth}p{0.27\textwidth}@{}}
\toprule
Authority & Construction & Reference unit & Question answered \\
\midrule
Expressed condition & One-hot elicitation label & One production instruction per clip & Was the requested acted condition recoverable from speech? \\
Recognized plurality & Most frequent of five votes; deterministic tie-break & One aggregated listener-derived class per clip & Which single class wins under the stated aggregation rule? \\
Recognized distribution & Seven-class vector of vote proportions & Complete five-vote histogram per clip & How should probability be allocated across the alternatives selected by listeners? \\
Evaluator-level view & Each recognition vote retained separately & One listener judgement, paired with the clip's expressed label & How much individual perception is hidden by clip-level aggregation? \\
\bottomrule
\end{tabular}
\end{table*}

Two analysis levels must therefore remain separate. At \emph{clip level}, each
utterance contributes once through its plurality or vote distribution. At
\emph{evaluator level}, every listener judgement contributes an
expressed--recognized pair before aggregation. Evaluator-level accuracy is
0.6638 over all votes, whereas clip-plurality accuracy is 0.7905. The difference
does not indicate a contradiction: five-vote aggregation often repairs an
individual disagreement. Evaluator level exposes the variation that this repair
compresses; it is not a fourth model target.

\section{Experimental Design}

\subsection{Frozen Representations and Probe Families}

The acoustic backbones are XLS-R 300M, HuBERT Large, and WavLM Large. Each has a
convolutional front end followed by 24 Transformer layers with hidden width
1,024. All backbone parameters remain frozen. A hidden state is therefore a
sequence of shape $T\times1024$, where $T$ depends on clip duration, and all
temporal means exclude padded frames.

The three probe families deliberately add limited trainable capacity. The
historically named \texttt{linear} probe mean-pools one selected hidden state,
projects 1,024 features to 128 with a ReLU, and maps the result to seven logits.
The \texttt{dense} probe first applies two frame-wise $1\times1$ convolutions
with 256 channels and ReLU activations, then mean-pools and classifies. It can
transform each frame non-linearly but does not model new temporal interactions.
The \texttt{agg\_dense} probe assigns a learned scalar to every one of the 25
available states--the convolutional state numbered 0 and Transformer layers
1--24--normalizes those scalars with a softmax, and sends the weighted sequence
through the dense head. Thus, the 25 points in its layer-weight plots are all
states made available to the aggregation probe; they are distinct from the ten
depths sampled by the single-layer probes.

Those sampled depths correspond to 0, 20, 33, 43, 53, 63, 73, 87, 90, and
100 per cent of the backbone, mapped to indices 0, 5, 8, 10, 13, 15, 18, 21,
22, and 24. For each target there are 60 single-layer conditions and three
all-layer conditions, giving 63 matched representation/probe configurations.
Across all three targets and five repeated runs, the design contains 945 probe
training trajectories.

\subsection{Splits and Training}

A fixed speaker-independent split assigns 4,142 clips to training, 1,380 to
validation, and 1,380 to testing. Every speaker occurs in one partition only,
and the same audio assignment is used for all authorities. Hard targets use
categorical cross-entropy; the recognized distribution uses soft
cross-entropy against the unsmoothed empirical vote vector. Probes are trained
for 30 epochs with Adam at a learning rate of $10^{-3}$. The retained checkpoint
maximizes validation accuracy for hard targets and minimizes validation soft
cross-entropy for the distributional target. Five runs vary probe
initialization and minibatch order while holding the split, features, and
settings fixed.

\begin{table}[t]
\caption{Experimental factors held in common across label authorities.}
\label{tab:design}
\centering
\scriptsize
\setlength{\tabcolsep}{4pt}
\begin{tabular}{@{}lp{0.58\columnwidth}@{}}
\toprule
Component & Setting \\
\midrule
Backbones & XLS-R 300M, HuBERT Large, WavLM Large; frozen \\
Single layers & 0, 5, 8, 10, 13, 15, 18, 21, 22, 24 \\
Probe families & \texttt{linear}, \texttt{dense}, \texttt{agg\_dense} \\
Split & Speaker-independent 60/20/20; 4,142/1,380/1,380 clips \\
Training & Adam, $10^{-3}$; 30 epochs; five runs \\
Selection & Validation checkpoint; reported layer/winner by mean test UAR \\
Grid & 63 conditions per target; 945 trajectories overall \\
\bottomrule
\end{tabular}
\end{table}

\subsection{Metrics and Uncertainty}

Hard decisions are assessed with unweighted average recall (UAR), the mean of
the seven class recalls. Unlike plain accuracy, UAR does not let a frequent
recognized class dominate the score. Agreement-stratified plots use ordinary
accuracy within each stratum because the question there is the probability of
matching the relevant hard label for clips with a given vote pattern.

Predicted probability vectors $q$ are compared with the observed vote
distribution $p$ using
\begin{equation}
D_{\mathrm{KL}}(p\Vert q)=\sum_c p_c\log\frac{p_c}{q_c},\qquad
\mathrm{Brier}(p,q)=\sum_c(q_c-p_c)^2.
\end{equation}
Both are proper distributional scores and lower is better. We also report the
top/tie-set hit rate as an intuitive secondary diagnostic, but do not use it to
choose between targets because the set becomes easier to hit as a tie expands.
Nominal Krippendorff's $\alpha$ describes listener reliability independently of
any model.

Means and sample standard deviations are computed over five aligned runs. For
headline winner contrasts, 1,000 paired clip-bootstrap replicates resample
clips while carrying all five predictions for a sampled clip together. The
resulting intervals quantify test-clip sampling conditional on the already
trained and selected probes. They do not include training or selection
uncertainty. Checkpoints are validation-selected, but the reported best layers
and individual winners are chosen post hoc by five-run mean test UAR. The 63
within-grid contrasts share data and representations and are therefore not
treated as 63 independent experimental units.

\section{Results}

The results are presented in the order needed to interpret the label-authority
question. We begin with the listeners, before asking any model to reproduce
their decisions. We then separate authority-dependent hard scores from
comparisons that hold the evaluation reference fixed, and finally examine what
the frozen representations contribute.

\subsection{Human Disagreement Before Model Scores}

\begin{figure*}[t]
\centering
\includegraphics[width=0.94\textwidth]{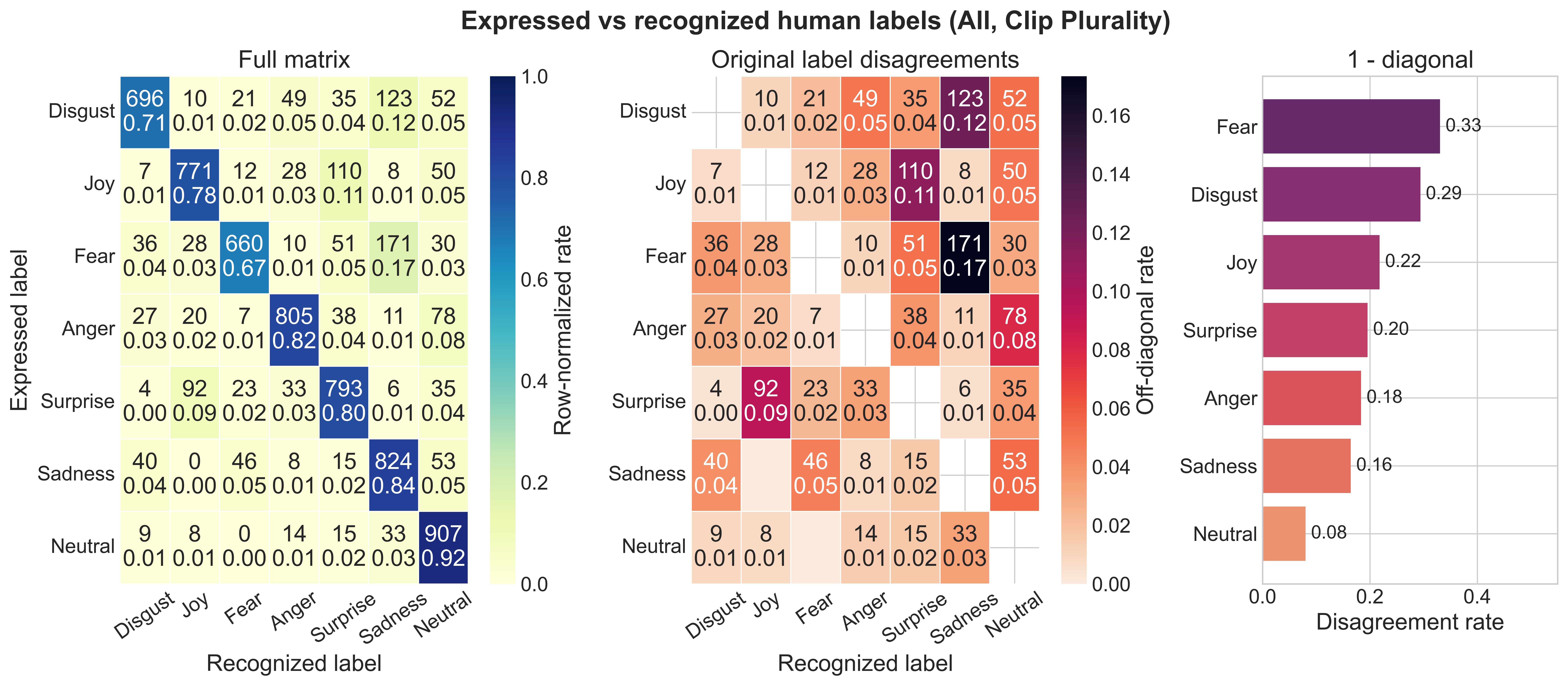}
\caption{Expressed condition versus the plurality of all five listener votes.
Rows are expressed labels and columns are recognized labels. The left panel
shows counts and row-normalized rates, the centre suppresses the diagonal, and
the right ranks expressed categories by their off-diagonal rate.}
\label{fig:human_confusion}
\end{figure*}

Figure~\ref{fig:human_confusion} compares the expressed condition with the
all-vote recognized plurality. The diagonal is strong: the plurality agrees
with the expressed label for 79.1\% of clips. It is not uniform, however.
Neutrality has a 0.92 row-normalized match rate, while fear reaches only 0.67.
The off-diagonal panel shows that disagreement has recurring semantic
directions rather than being spread evenly across all categories.

The aggregation level changes what ``human accuracy'' means. At evaluator
level, every judgement is counted separately and 66.4\% match the expressed
condition. At clip level, five judgements are first collapsed to one plurality,
raising the match rate to 79.1\%. The latter is a consensus statistic, not an
estimate of how often an individual listener recognizes the intended category.
This distinction explains why the corpus-level plurality appears more stable
than its constituent votes.

The three leading undirected pairs--fear/sadness, joy/surprise, and
disgust/sadness--also remain prominent whether each clip or each evaluator is
the unit of analysis. At clip level they account for 217, 202, and 163
mismatches, respectively. Plurality compression reduces the amount of
disagreement without replacing its main categorical structure.

\begin{table*}[t]
\caption{Human-label summaries. Panel (a) distinguishes individual judgements
from clip pluralities and reports nominal Krippendorff's $\alpha$. Panel (b)
partitions all 6,902 clips into mutually exclusive agreement strata.}
\label{tab:human_summary}
\centering
\scriptsize
\begin{minipage}[t]{0.48\textwidth}
\centering
\textbf{(a) Analysis level and reliability}\\[0.3em]
\setlength{\tabcolsep}{3.5pt}
\begin{tabular}{@{}lrrrr@{}}
\toprule
Votes & Ratings & Eval. acc. & Clip acc. & $\alpha$ \\
\midrule
All & 34,510 & 0.6638 & 0.7905 & 0.476 \\
Good quality & 33,200 & 0.6792 & 0.7937 & 0.501 \\
\bottomrule
\end{tabular}
\end{minipage}\hfill
\begin{minipage}[t]{0.48\textwidth}
\centering
\textbf{(b) Agreement structure}\\[0.3em]
\setlength{\tabcolsep}{4pt}
\begin{tabular}{@{}lrr@{}}
\toprule
Stratum & Clips & Plurality $=$ expressed \\
\midrule
Unanimous (share 1.00) & 1,792 & 1.000 \\
Share 0.80 & 1,909 & 0.870 \\
Share 0.60 & 2,070 & 0.663 \\
Tied plurality & 715 & 0.603 \\
Unique, share 0.40 & 416 & 0.481 \\
\bottomrule
\end{tabular}
\end{minipage}
\end{table*}

Table~\ref{tab:human_summary}(b) supplies an important boundary condition. On
the 1,792 unanimous clips, the five listeners always converge on the expressed
condition. As votes disperse, agreement between authorities declines steadily;
when the winning class has only two of five votes without a tie, it falls below
one half. A hard recognized label is consequently most contestable on exactly
the clips where an aggregation rule is doing the most work. Removing ratings
flagged for lower audio quality changes $\alpha$ only from 0.476 to 0.501, so
audio filtering does not explain away the disagreement.

\subsection{What the Hard-Label Gap Means}

Table~\ref{tab:best_uar} reports the target-specific winners. The best expressed
probe reaches $0.825\pm0.009$ UAR, compared with $0.746\pm0.006$ for the best
hard-recognized probe. Their aligned-run difference is $0.079\pm0.014$; a paired
clip bootstrap gives a conditional 95\% interval of $[0.058,0.099]$. This is a
real descriptive separation on the present test split, but it compares two
questions: the reference changes from an elicited condition to a plurality of
listener judgements, and the selected layer also changes.

\begin{table}[t]
\caption{Target-specific best own-reference hard UAR. Expressed models use the
expressed reference; recognized models use plurality, with the soft target read
by argmax. Values are mean $\pm$ sample SD over five runs.}
\label{tab:best_uar}
\centering
\scriptsize
\begin{tabular}{@{}llcc@{}}
\toprule
Training target & Best configuration & Layer & UAR \\
\midrule
Expressed & XLS-R Dense & 13 & $0.825\pm0.009$ \\
Recognized plurality & XLS-R Dense & 15 & $0.746\pm0.006$ \\
Recognized distribution & WavLM Dense & 10 & $0.749\pm0.004$ \\
\bottomrule
\end{tabular}
\end{table}

\begin{table}[t]
\caption{Common-reference evaluation, mean $\pm$ sample SD over five runs
after equally averaging nine target-selected backbone/probe configurations.
Hard UAR uses recognized plurality; KL and Brier use the five-vote histogram.}
\label{tab:common_reference}
\centering
\scriptsize
\begin{tabular}{@{}lccc@{}}
\toprule
Training target & Hard UAR & KL & Brier \\
\midrule
Expressed & $0.739\pm0.002$ & $1.667\pm0.156$ & $0.272\pm0.009$ \\
Rec. plurality & $0.731\pm0.005$ & $1.223\pm0.145$ & $0.236\pm0.008$ \\
Rec. distribution & $0.736\pm0.005$ & $\mathbf{0.510\pm0.002}$ & $\mathbf{0.162\pm0.001}$ \\
\bottomrule
\end{tabular}
\end{table}

\begin{figure*}[t]
\centering
\begin{minipage}[t]{.42\textwidth}
  \centering
  \includegraphics[width=\linewidth]{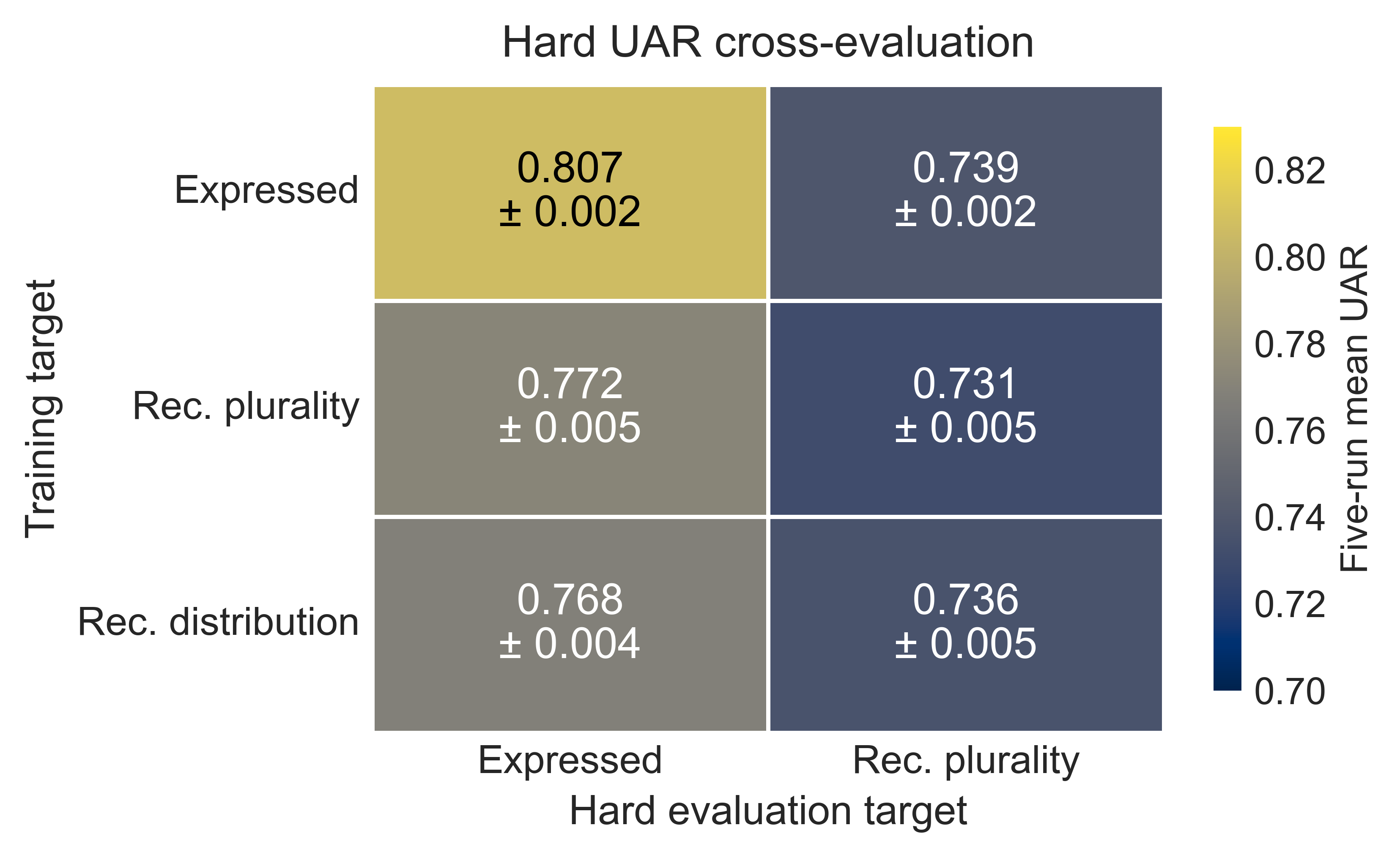}\\[-.2em]
  \footnotesize (a)
\end{minipage}\hfill
\begin{minipage}[t]{.53\textwidth}
  \centering
  \includegraphics[width=\linewidth]{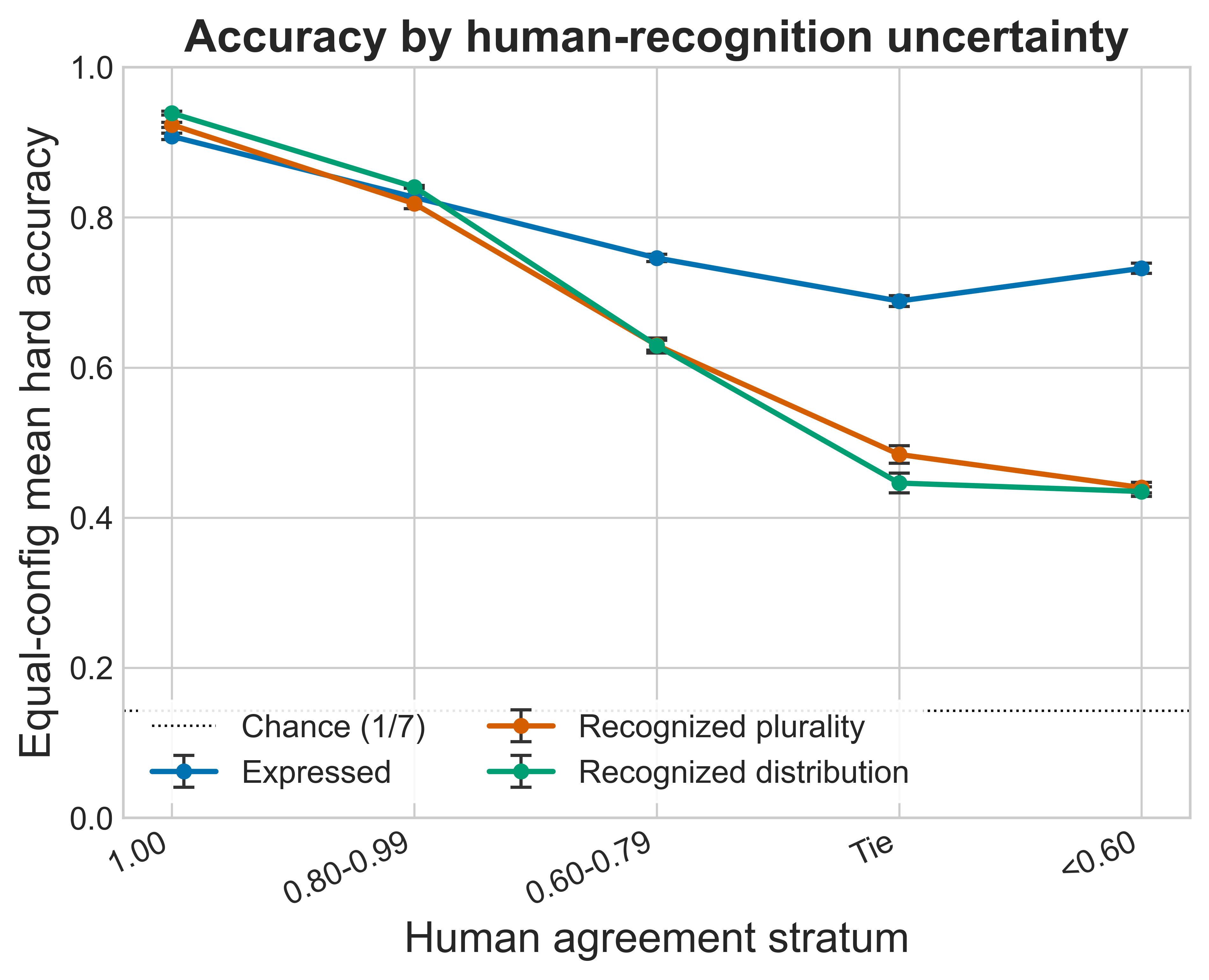}\\[-.2em]
  \footnotesize (b)
\end{minipage}
\caption{Authority and agreement diagnostics. (a) Hard-UAR cross-evaluation:
rows are training authorities and columns are hard evaluation authorities.
Cells average nine target-selected configurations within each run and report
mean $\pm$ sample SD over five runs. (b) Hard accuracy by mutually exclusive
listener-agreement stratum. Expressed probes use the expressed reference;
recognized probes use plurality.}
\label{fig:hard_and_agreement}
\end{figure*}

Figure~\ref{fig:hard_and_agreement}(a) holds the evaluation authority fixed and
therefore answers a different question. Against expressed labels, expressed
training performs best; against listener plurality, the three rows are much
closer and none establishes a general advantage across both columns. The same
models retain information about both authorities, but supervision aligns their
decisions with different references. As secondary consistency evidence, all 63
matched recognized-minus-expressed own-reference deltas are negative (mean
$-0.065$, range $-0.095$ to $-0.028$). These dependent conditions demonstrate
directional consistency, not 63 independent replications.

Agreement strength explains where the hard recognized task becomes difficult.
In Fig.~\ref{fig:hard_and_agreement}(b), hard-plurality training falls from
$0.923\pm0.003$ accuracy on 381 unanimous test clips to $0.484\pm0.012$ on 145
ties and $0.440\pm0.007$ on 82 clips with a unique winner below 0.60 vote share.
The soft-trained model follows the same hard-argmax pattern. Expressed accuracy
uses another reference, so its higher values in the low-agreement strata do not
show that the production condition is a better description of perception.
They show that plurality difficulty is coupled to the vote pattern from which
the label was constructed.

\subsection{Do Model Errors Resemble Human Errors?}

The most frequent model confusions are not arbitrary. Joy/surprise,
fear/sadness, and disgust/sadness lead under all three targets, matching the
human ranking described above. To broaden the comparison beyond those examples,
Table~\ref{tab:model_human_overlap} compares each selected
configuration's ten most frequent undirected error pairs with the fixed
evaluator-level human top ten. At least six pairs overlap for every selected
configuration and run.

\begin{table}[t]
\caption{Top-10 model--human confusion-pair overlap. The mean averages nine
selected backbone/probe configurations within each run; the SD is over five
runs, and the range is across configuration means.}
\label{tab:model_human_overlap}
\centering
\scriptsize
\setlength{\tabcolsep}{4pt}
\begin{tabular}{@{}lcc@{}}
\toprule
Training target & Mean $\pm$ SD & Config. range \\
\midrule
Expressed & $0.660\pm0.013$ & $[0.600,0.700]$ \\
Recognized plurality & $0.707\pm0.017$ & $[0.680,0.740]$ \\
Recognized distribution & $\mathbf{0.733\pm0.014}$ & $[0.680,0.780]$ \\
\bottomrule
\end{tabular}
\end{table}

Listener-derived supervision is descriptively closer to the human ranking, with
the soft target highest at $0.733\pm0.014$. This rank-based diagnostic does not
measure calibrated probabilities or establish a statistically significant
advantage. Its narrower message is nevertheless useful: when recognized-target
models fail, they often fail at boundaries on which listeners themselves are
divided. The lower hard score is partly inherited from the ambiguity of the
reference rather than produced by an unrelated model error structure.

\subsection{What Distributional Supervision Recovers}

Table~\ref{tab:common_reference} evaluates all targets against the same
listener-derived references. Hard and soft recognized training have close hard
plurality UAR, $0.731\pm0.005$ and $0.736\pm0.005$. Against the complete vote
histogram, however, soft supervision reduces KL from $1.223\pm0.145$ to
$0.510\pm0.002$ and Brier from $0.236\pm0.008$ to $0.162\pm0.001$. The hard
winner changes little while the allocation of probability among alternatives
changes substantially.

\begin{figure}[t]
\centering
\begin{minipage}[t]{.48\columnwidth}
  \centering
  \includegraphics[width=\linewidth]{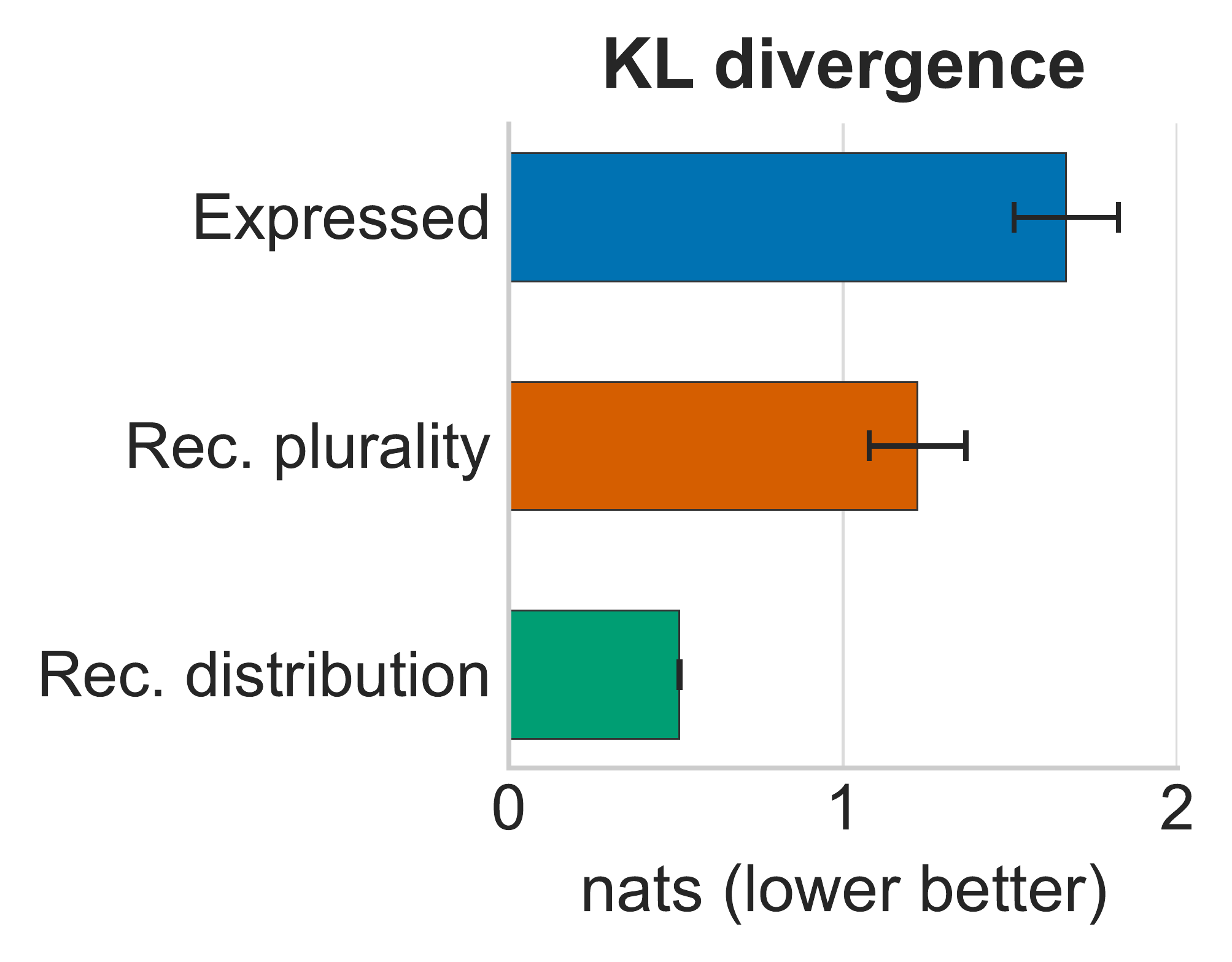}\\[-.2em]
  \footnotesize (a)
\end{minipage}\hfill
\begin{minipage}[t]{.48\columnwidth}
  \centering
  \includegraphics[width=\linewidth]{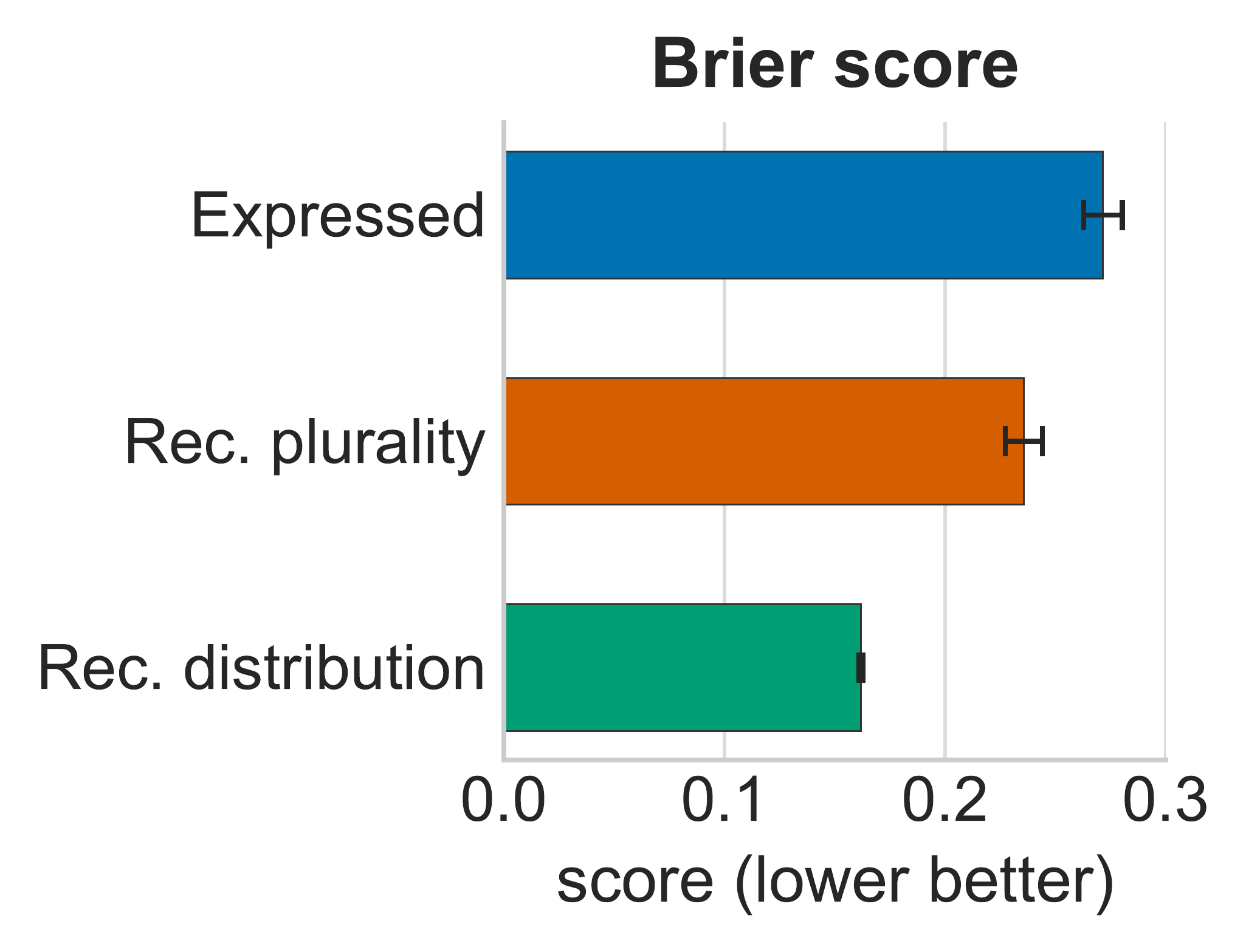}\\[-.2em]
  \footnotesize (b)
\end{minipage}
\caption{Alignment with the observed five-vote distribution: (a) KL divergence
and (b) multiclass Brier score. Bars equally average nine target-selected
configurations within each run; error bars are sample SDs over five runs. Lower
is better.}
\label{fig:distributional_alignment}
\end{figure}

Figure~\ref{fig:distributional_alignment} shows that the aggregate advantage is
not driven by one favourable backbone. Soft training has the lowest mean Brier
score in all nine matched backbone/probe comparisons. A
configuration-matched sensitivity, which anchors both recognized targets to
the hard target's representation condition, preserves the ordering: soft minus
hard is $-0.713\pm0.144$ for KL and $-0.074\pm0.009$ for Brier, while hard UAR
changes by only $+0.002\pm0.007$.

\begin{figure*}[t]
\centering
\includegraphics[width=0.82\textwidth]{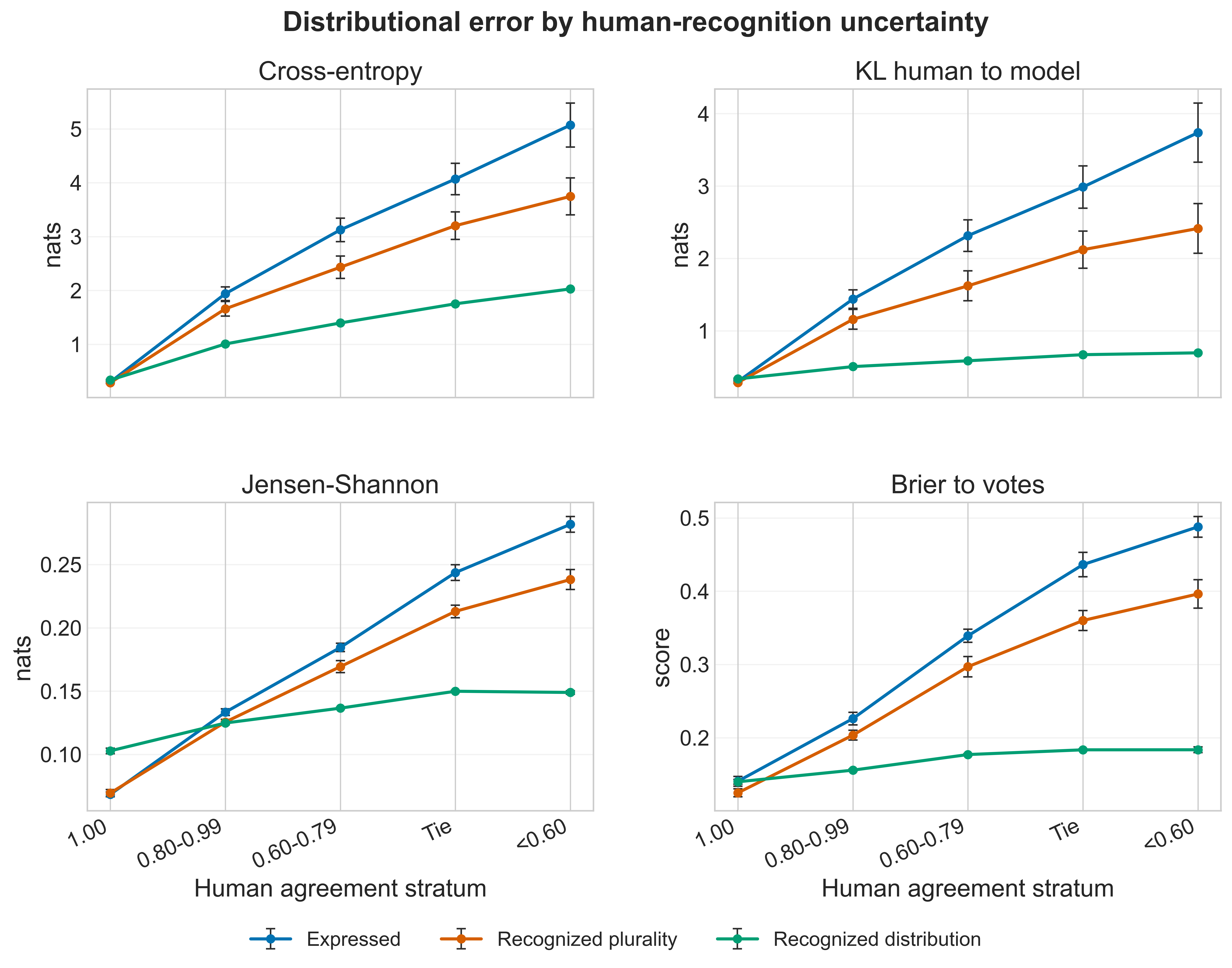}
\caption{Distributional error against the observed vote histogram by human
agreement stratum. Panels show cross-entropy, KL divergence, Jensen--Shannon
divergence, and Brier score. Each point averages nine selected configurations
within each run; error bars are sample SDs over five runs. KL and Brier provide
the primary proper-score evidence; the other panels are supporting diagnostics.}
\label{fig:distribution_by_agreement}
\end{figure*}

Figure~\ref{fig:distribution_by_agreement} localizes the difference. All models
find diffuse vote histograms harder to match, but soft training remains closest
to those histograms as agreement weakens. The apparent paradox on ties is
especially revealing. A matched soft-trained probe can assign probability to
both co-winning classes and improve KL and Brier, yet lose hard UAR because the
metric credits only the class selected by the deterministic tie-break.

\begin{table}[t]
\caption{Shared-reference results on ambiguous test subsets. Values are mean
$\pm$ sample SD over five runs after equally averaging nine selected
configurations. Lower KL and Brier are better.}
\label{tab:ambiguous_subsets}
\centering
\scriptsize
\setlength{\tabcolsep}{3pt}
\begin{tabular}{@{}llccc@{}}
\toprule
Subset & Training & Hard UAR & KL & Brier \\
\midrule
Tie ($n=145$) & Hard & $0.469\pm0.015$ & $2.121\pm0.256$ & $0.360\pm0.014$ \\
 & Soft & $0.435\pm0.012$ & $\mathbf{0.671\pm0.004}$ & $\mathbf{0.184\pm0.001}$ \\
$<0.60$ ($n=82$) & Hard & $0.443\pm0.010$ & $2.415\pm0.343$ & $0.396\pm0.019$ \\
 & Soft & $0.439\pm0.006$ & $\mathbf{0.696\pm0.008}$ & $\mathbf{0.184\pm0.004}$ \\
\bottomrule
\end{tabular}
\end{table}

Table~\ref{tab:ambiguous_subsets} quantifies that trade-off. On ties, soft
training gives up hard UAR against the arbitrary winner while sharply reducing
both proper scores. Below 0.60 vote share, hard UAR is nearly unchanged and the
distributional improvement remains. For best individual models, paired
clip-bootstrap intervals exclude zero for soft-minus-hard KL and Brier both on
ties and below 0.60. They remain conditional on trained, test-selected winners.
Re-scoring against good-quality votes preserves the same ordering without
retraining.

\subsection{What the Layer Mixtures Add}
\begin{figure*}[t]
\centering
\includegraphics[width=0.98\textwidth]{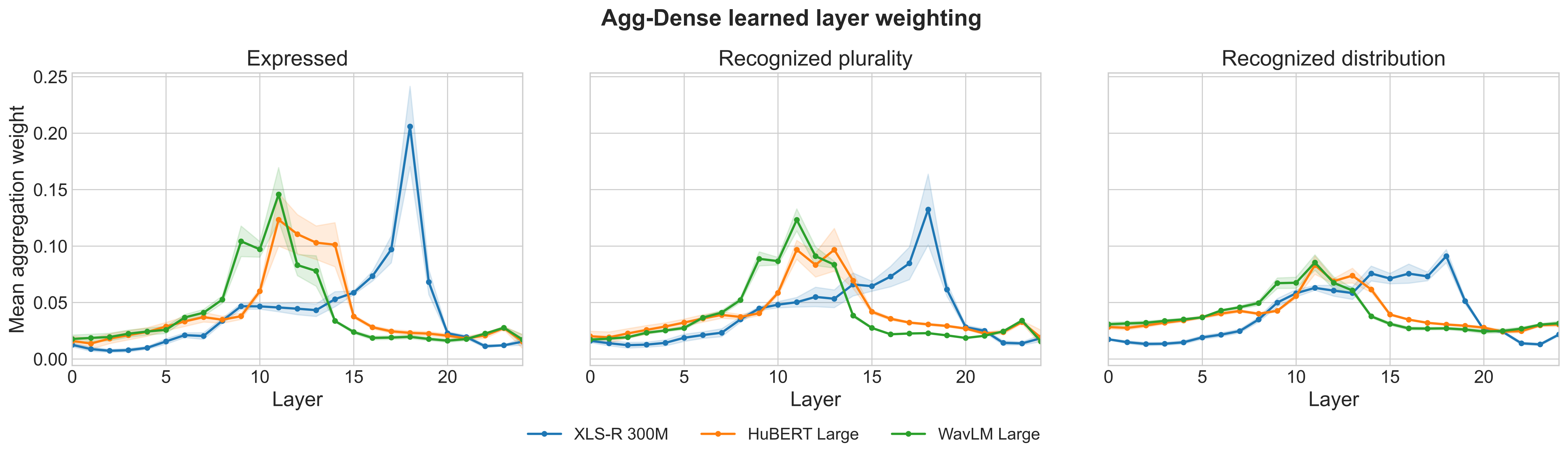}
\caption{Learned \texttt{agg\_dense} mixture weights for expressed,
recognized-plurality, and recognized-distribution supervision. Lines are
five-run means and bands are sample SDs. Layer 0 is the convolutional state;
layers 1--24 are Transformer states.}
\label{fig:agg_weights}
\end{figure*}

The \texttt{agg\_dense} probe asks whether a learned mixture of all frozen states
is preferable to selecting one sampled state. Figure~\ref{fig:agg_weights}
shows all 25 coefficients--state 0 from the convolutional front end and states
1--24 from the Transformer--rather than the ten depths used by the
single-layer grid. The profiles generally favour middle-to-late states, but
they also change with the target. For example, the expressed condition produces
a sharper XLS-R peak near layer 18 than the listener-distribution target. The
weights are fitted associations, not causal importance scores for individual
layers.

\begin{table*}[t]
\caption{Best single-layer dense probe versus \texttt{agg\_dense}. Values are
hard UAR mean $\pm$ sample SD over five aligned runs; $\Delta$ is aggregation
minus single layer. Soft-target models are evaluated against hard plurality.}
\label{tab:aggregation}
\centering
\scriptsize
\setlength{\tabcolsep}{4pt}
\begin{tabular}{@{}llcccc@{}}
\toprule
Target & Backbone & Best single layer & Single-layer UAR & Aggregate UAR & Paired $\Delta$ \\
\midrule
Expressed & XLS-R & Dense L13 & $0.825\pm0.010$ & $0.801\pm0.006$ & $-0.023\pm0.013$ \\
 & HuBERT & Dense L13 & $0.808\pm0.007$ & $0.796\pm0.008$ & $-0.012\pm0.014$ \\
 & WavLM & Dense L18 & $0.807\pm0.004$ & $0.822\pm0.010$ & $+0.015\pm0.010$ \\
\addlinespace
Recognized plurality & XLS-R & Dense L15 & $0.746\pm0.006$ & $0.719\pm0.016$ & $-0.026\pm0.012$ \\
 & HuBERT & Dense L13 & $0.729\pm0.007$ & $0.721\pm0.011$ & $-0.008\pm0.009$ \\
 & WavLM & Dense L10 & $0.739\pm0.007$ & $0.739\pm0.009$ & $\approx0\pm0.013$ \\
\addlinespace
Recognized distribution & XLS-R & Dense L18 & $0.748\pm0.013$ & $0.743\pm0.024$ & $-0.005\pm0.020$ \\
 & HuBERT & Dense L10 & $0.731\pm0.009$ & $0.732\pm0.010$ & $+0.000\pm0.006$ \\
 & WavLM & Dense L10 & $0.749\pm0.004$ & $0.734\pm0.014$ & $-0.016\pm0.014$ \\
\bottomrule
\end{tabular}
\end{table*}

Table~\ref{tab:aggregation} shows no uniform hard-UAR advantage for aggregation.
It helps WavLM on the expressed target, is approximately level in several
conditions, and trails the post-hoc best single layer elsewhere. The comparator
is optimistic because it was chosen from the same test split, while aggregation
uses all states and a different effective input. The appropriate conclusion is
therefore modest: representation depth interacts with the chosen authority,
but a learned mixture is not a universally better substitute for layer
selection.

\subsection{Uncertainty of the Headline Contrasts}

The uncertainty summaries answer complementary questions. Run SDs describe
variation from fitting the probe repeatedly. The paired clip bootstrap describes
which test clips happened to be sampled, conditional on the selected models.
Table~\ref{tab:bootstrap} reports the four full-test contrasts that anchor the
paper's conclusions.

\begin{table*}[t]
\caption{Best-individual paired contrasts on all 1,380 test clips. The first
column gives the mean $\pm$ sample SD of five aligned run differences; the
interval comes from 1,000 paired clip-bootstrap replicates and excludes
training and model-selection uncertainty. Each contrast is left minus right.}
\label{tab:bootstrap}
\centering
\scriptsize
\setlength{\tabcolsep}{6pt}
\begin{tabular}{@{}lcc@{}}
\toprule
Comparison & Paired-run $\Delta$ & Conditional 95\% CI \\
\midrule
Expressed $-$ hard recognized, own-reference UAR & $+0.079\pm0.014$ & $[0.058,0.099]$ \\
Soft $-$ hard recognized, recognized UAR & $+0.003\pm0.008$ & $[-0.013,0.019]$ \\
Soft $-$ hard recognized, KL & $-0.360\pm0.081$ & $[-0.399,-0.325]$ \\
Soft $-$ hard recognized, Brier & $-0.056\pm0.015$ & $[-0.066,-0.046]$ \\
\bottomrule
\end{tabular}
\end{table*}

The hard soft-minus-hard interval crosses zero. This neither establishes a hard
accuracy advantage nor proves equivalence. The proper-score intervals exclude
zero in the direction favouring soft training. By contrast, the larger
expressed-minus-recognized interval still compares different references and
must not be interpreted as a causal penalty from changing only the training
target.

\section{Discussion}

\subsection{Ground Truth as a Task Specification}

The three authorities are related, but they are not interchangeable versions
of the same label. The expressed condition identifies what the speaker was
asked to perform. The recognized plurality identifies the class that wins after
five perceptions are compressed by an aggregation and tie rule. The recognized
distribution retains the alternatives chosen by those listeners. Each can be a
legitimate target, provided that the system's claim is phrased in terms of the
question that target can answer.

This distinction changes the meaning of error. An expressed-label error misses
the elicited condition. A plurality error misses the panel's operational winner,
which may itself depend on a tie-break. A distributional error places
probability differently from the observed pattern of votes. Treating these
events as one generic failure obscures the main result. The own-reference hard
UAR gap does not reveal a universal reduction in SER capability; it shows that
recovering the performed condition and predicting the perceptual winner are
different tasks. Cross-evaluation confirms that the representations retain
information about both, while the agreement analysis explains why one hard
reference becomes less stable on ambiguous clips.

The distributional result supplies a constructive response to that instability.
On the shared recognized reference, soft and hard training produce similar hard
UAR over the complete test split. The soft model is nevertheless much closer to
the full vote distribution under proper scores. These statements are compatible
because an argmax asks only which class is largest, whereas KL and Brier ask how
the whole probability vector is arranged. Tied clips make the distinction
visible: a model can respect both co-winning emotions and be penalized for not
guessing the one selected by a deterministic rule.

Human--model confusion alignment adds an interpretive layer. The models do not
simply become noisy when listener labels are introduced. Their most frequent
errors remain concentrated around fear/sadness, joy/surprise, and
disgust/sadness, the same boundaries that organize listener disagreement. This
does not make every model error a human-like perception, nor does a top-10
overlap establish calibration. It does show that much of the observed
difficulty is structured by the perceptual problem itself.

\subsection{Implications for Internet of Sounds Systems}

For a connected audio system, label authority is part of the system
specification rather than a detail of dataset preparation. A production archive
may legitimately index the intended affect of recorded material. A user-facing
search engine may instead need to retrieve clips by likely perceived affect. An
accessibility service or affect-aware interface may need to expose uncertainty
when listeners would disagree, because an overconfident adaptation can be more
harmful than a deferred decision. The same audio model cannot be evaluated for
all three purposes by changing only the prose around one score.

This has consequences at the dataset and service boundary. Production
condition, vote counts, annotator count, aggregation rule, and quality metadata
should remain separate fields rather than being overwritten by one
``ground-truth'' class. A service can still derive a categorical label when a
routing or retrieval interface requires it. Preserving provenance allows that
decision to be audited, recalculated under another rule, or accompanied by an
agreement threshold. It also allows a downstream component to request more
annotations or abstain when the perceptual evidence is diffuse.

The present study does not evaluate networking, latency, or deployment. Its
Internet of Sounds contribution is upstream: it clarifies what an affective
audio component is being asked to predict and which evidence should accompany
that prediction before the component is embedded in a connected application.
Testing whether the proposed metadata and uncertainty signals improve a
deployed interaction remains future work.

\subsection{A Reporting Protocol for Label Authority}

Three reporting practices follow from the evidence. First, an expressed-label
score should be described as recovery of an elicited production condition, not
automatically as perceived-emotion accuracy. Second, when one listener-derived
class is required, the aggregation rule, tie policy, panel size, and agreement
strata should be reported with the hard score. Third, when perceptual ambiguity
is itself the target, the vote distribution should be retained and evaluated
with a proper score; any hard decision should be reported separately.

These practices also prevent unlike metrics from being used to rank unlike
authorities. The best-model expressed--recognized gap changes both the reference
and selected model. By contrast, the hard and soft recognized rows in
Table~\ref{tab:common_reference} share the hard-plurality and vote-distribution
references. Claims about the effect of a training target should hold the
evaluation authority and, where possible, the representation and probe fixed.
Claims about authority should instead acknowledge that the estimand itself has
changed.

\section{Limitations and Outlook}

The evidence comes from one corpus, one language, simulated emotional speech,
and five votes per clip. Another listener panel could change a weak plurality,
and cultural or contextual differences could change the confusion structure.
The good-quality sensitivity check shows that filtering lower-quality ratings
does not remove the main pattern, but it cannot separate cultural, contextual,
and individual sources of disagreement. Follow-up annotation should retain
evaluator identities and collect enough judgements to estimate both rater- and
clip-level variation.

The observed vote histograms are finite samples rather than latent perceptual
distributions. KL and Brier therefore contain sampling variation even for a
well-specified model. Subtracting a shared finite-vote reference level would
preserve between-target differences on the same clips, but it would not turn a
raw score into pure model error or justify unqualified comparisons between
agreement strata. A larger or repeated annotation panel is needed to study how
stable the inferred distributions are.

The modelling scope is similarly bounded. Frozen backbones and shallow probes
measure what is accessible without updating the speech representation; they do
not establish what end-to-end fine-tuning can learn. Checkpoints use validation
data, but the best layers and individual winners were chosen by mean test UAR.
Winner-based results are consequently descriptive and optimistic. The paired
bootstrap conditions on those fitted and selected models. A confirmatory study
should predeclare its contrasts, select all hyperparameters on validation data,
and reserve an untouched test set for interval estimation.

Cross-corpus validation is also necessary. Grossi et al. report limited transfer
from Emozionalmente intended-label training to external speech subsets
\cite{grossi2024actors}. That result motivates replication but cannot isolate
label authority because speakers, acoustics, class support, and annotation all
change together. A stronger extension would collect or align production labels,
pluralities, and vote distributions across languages and across acted and
spontaneous speech under one protocol. Fine-tuned and frozen models could then
be compared without changing the measurement question.

The layer-mixture analysis should also remain exploratory. The target-dependent
profiles suggest that different authorities draw on frozen representations in
different ways, but the weights are correlated fitted coefficients rather than
causal attributions. Aggregation and single-layer probes also differ in their
inputs and selection procedures. These observations motivate a
validation-selected representation study; they do not identify one universally
optimal layer or backbone.

\section{Conclusion}

Within Emozionalmente, label authority changes both what an SER system is asked
to predict and how its output should be evaluated. Expressed labels provide a
controlled account of production. Listener pluralities provide an operational
single-class decision, but become brittle as agreement weakens. Vote
distributions preserve the perceptual alternatives that a hard plurality
discards, and distributional supervision represents those alternatives more
faithfully even when hard UAR changes little.

No authority therefore replaces every other ground truth. The practical lesson
is to make the authority visible: state what generated the label, preserve how
listener evidence was aggregated, report agreement when a hard perceptual class
is used, and score probability distributions when disagreement belongs to the
target. Within the limits of this corpus and probing design, that discipline
makes SER conclusions more transparent and more useful for human-facing
Internet of Sounds applications. Code will be released by the conference.

\IEEEtriggeratref{10}
\bibliographystyle{IEEEtran}
\bibliography{references}

\end{document}